# Asymptotic Regge Trajectories of Non-strange Mesons

K. A. Bugaev[1], E. G. Nikonov[2], A. S. Sorin[3] and G. M. Zinovjev[1]

[1]Bogolyubov Institute for Theoretical Physics, National Academy of Sciences of Ukraine,
[2]Laboratory for Information Technologies, JINR, Dubna, Russia
[3]Bogoliubov Laboratory of Theoretical Physics, JINR, Dubna, Russia

**Abstract**

We analyze the asymptotic behavior of Regge trajectories of non-strange mesons. In contrast to an existing belief, it is demonstrated that for the asymptotically linear Regge trajectories the width of heavy hadrons cannot linearly depend on their mass. Using the data on masses and widths of $\rho_{J^{--}}$, $\omega_{J^{--}}$, $a_{J^{++}}$ and $f_{J^{++}}$ mesons for the spin values $J \leq 6$, we extract the parameters of the asymptotically linear Regge trajectory predicted by the finite width model of quark gluon bags. As it is shown the obtained parameters for the data set B correspond to the cross-over temperature lying in the interval 170.9–175.3 MeV which is consistent with the kinetic freeze-out temperature of early hadronizing particles found in relativistic heavy ion collisions at and above the highest SPS energy.

**Key words:** Linear and nonlinear Regge trajectories, resonance mass, resonance width

**PACS:** 25.75.Nq, 25.75.-q

## 1 Introduction

Regge poles method has been introduced in particle physics at the beginning of 60-ties [1, 2] and since that time it is widely used to describe the high-energy interactions of hadrons and nuclei. This method establishes an important connection between high energy scattering and spectrum of particles and resonances. Also it was a starting point to develop the dual and string models of hadrons. Although the apex of Regge method in particle physics ended after the beginning of quantum chromodynamics (QCD) era, till present days it serves as a reliable tool to describe a variety of non-perturbative QCD processes.

A Regge trajectory $J = \alpha(s)$ is expressed commonly in terms of the center-of-mass energy squared $s$ of colliding particles. Due to a crossing symmetry of strong interaction it is often used in $t$ or $u$ channels in terms of the Mandelstam variables $t$ or $u$, respectively. $\alpha(s)$ represents a set of leading Regge poles on the complex plane of angular momentum. An astonishing (approximate) linearity of Regge trajectories for the known hadronic states of mass $M_h$ and spin $J_h$, i.e. $J_h = \alpha(M_h^2) = \alpha_0 + \alpha_0' M_h^2$, remains one of the unresolved problems of QCD despite some promising results obtained within the phenomenological planar models [3, 4]. Up to now there is no consensus [5] on a linearity of established Regge trajectories of hadrons due to the lack of the experimental data on heavy resonances having spin above 6. However, it is necessary to mention that in 60-ies and 70-ies of last century a lot of efforts was put into study of the asymptotic behavior of Regge trajectories for $|s| \to \infty$ using their analytical properties in complex $s$-plane [6, 7]. Under some general assumptions (see later) it was found [8] that asymptotic Regge trajectories cannot grow faster than the linear function of $s$ and cannot increase slower than the square root of $s$ for $|s| \to \infty$. This general result restricts the non-linear behavior from above, but it does not allow to rule out the nonlinear $s$-behavior that is weaker than 1 and stronger than $\frac{1}{2}$.

Unexpectedly, a strong argument in favor of linear asymptotic behavior of Regge trajectories of free hadrons was given in [9] by the exactly solvable statistical model for quark gluon (QG) bags [10]. This model is named the finite width model (FWM) since it accounts for finite width of QG bags. Quite generally, the FWM demonstrates that free QG bags of mass $M_r \geq M_0 \approx 2.5$ GeV that belong to the continuous mass-volume spectrum of the Hagedorn type [11] should have the mean width $\Gamma_r \sim \sqrt{M_r}$. Such a behavior can be provided by the asymptotically linear Regge trajectory [8, 9] only. Moreover, the FWM shows that such a behavior of mean width of QG bags, but with the temperature dependent coefficient, remains valid at high temperatures.

The main purpose of this work is to extract the parameters of the asymptotic Regge trajectory predicted by the FWM using the experimental data. For our analysis we take the data of $\rho_{J^{--}}$, $\omega_{J^{--}}$, $a_{J^{++}}$ and $f_{J^{++}}$ mesons which are known with the highest accuracy for $J \leq 6$. In contrast to the usual analysis which is restricted by searching for a connection between spin and mass of the trajectory members, we use the full Regge trajectory in the complex $s$-plane. This allows us to simultaneously describe the masses and widths of involved mesons. Such a task is very important nowadays since a great significance of the width of heavy resonances or bags for the realistic equation of state of strongly interacting matter [9, 10, 12] and for a description of the fast equilibration process of heavy baryons/antibaryons [13, 14, 15] and kaons/antikaons [15] in relativistic heavy ion collisions was realized only recently. Also our analysis is necessary for further development of the string model of hadrons and for improvement of such transport codes as the hadron string dynamics [16] and the UrQMD model [17] by including the finite width of heavy resonances. In addition here we would like to demonstrate that a widely spread belief [14, 15, 18] that the width of heavy hadronic resonances linearly depends on their mass is simply inconsistent with an existence of linear Regge trajectories.

The work is organized as follows. Section 2 contains a brief analysis of the asymptotic Regge trajectory properties which demonstrates that the resonance width is proportional to its mass for the asymptotically nonlinear trajectories only. In Section 3 we discuss two hypotheses to be verified and define the corresponding data sets to be fitted. The details of the fitting procedure and the discussion of the obtained results are given in Section 4, while Section 5 contains our conclusions.

## 2 Asymptotic Behavior of Bosonic Regge Trajectories

In the pre-QCD era a lot of efforts was put forward [6, 8] to determine the asymptotic behavior of Regge trajectories of hadronic resonances. In our analysis we follow Ref. [8] and adopt the most general assumptions on the trajectory: (I) $\alpha(s)$ is an analytical function, having only the physical cut from $s = s_0$ to $s = \infty$; (II) $\alpha(s)$ is polynomially restricted at the whole physical sheet; (III) there exists a finite limit of the phase trajectory at $s \to \infty$. Using these assumptions, it was possible to prove [8] that for $s \to \infty$ the asymptotic behavior of Regge trajectory at the whole physical sheet is

$$\alpha_u(s) = -g^2 \left[-s\right]^\nu, \quad \text{with} \quad \frac{1}{2} \leq \nu \leq 1. \quad (1)$$

Here the function $g^2 > 0$ should increase slower than any power in this limit and its phase must vanish at $|s| \to \infty$. Clearly, $\nu = 1$ is an upper bound for the asymptotic behavior while $\nu = \frac{1}{2}$ is its lower bound.

Since our main interest here is related to the asymptotically linear trajectories, we restrict ourselves to the trajectories of the form

$$\alpha(s) = g^2 \left[-(-s)^\nu + q(s)\right], \quad \text{with} \quad \frac{1}{2} \leq \nu \leq 1, \quad (2)$$

where $g$ is real constant and the correction $q(s)$ increases slower than $|s|^\nu$ in the limit $|s| \to \infty$, i.e. $|q(s)|/|s|^\nu \to 0$ in this limit. Since at the resonance position $s = s_r = |s_r| e^{i\phi_r}$ in the complex $s$-plane the trajectory defines its spin $J_r$, one obtains $\text{Im}\,[\alpha(s_r)] = 0$. This condition allows one to determine the phase of physical trajectory from the equation

$$\sin(\nu\,\phi_r + \pi(1-\nu)) = -\text{Im}[q(s_r)]|s_r|^{-\nu} \to 0, \tag{3}$$

which has a formal solution $\phi_r = \pi\left(1 - \frac{1}{\nu}\right) - \frac{1}{\nu}\text{Im}[q(s_r)]|s_r|^{-\nu}$. Considering the complex energy plane $E = \sqrt{s_r} \equiv M_r - i\frac{\Gamma_r}{2}$, one can determine the mass $M_r$ and the width $\Gamma_r$

$$M_r = |s_r|^{\frac{1}{2}} \cos\tfrac{\phi_r}{2} \quad \text{and} \quad \Gamma_r = -2\,|s_r|^{\frac{1}{2}} \sin\tfrac{\phi_r}{2} \tag{4}$$

of a resonance belonging to the trajectory (2). It is clear that positive values of resonance mass and width correspond to the inequalities $-\frac{\pi}{2} < \frac{\phi_r}{2} < 0$ which in the limit $|s_r| \to \infty$ lead to the following conditions

$$-1 < \left(1 - \tfrac{1}{\nu}\right) < 0 \quad \Rightarrow \quad \tfrac{1}{2} < \nu < 1. \tag{5}$$

For the limiting cases $\nu = 1$ and $\nu = \frac{1}{2}$ the positive value of width and mass, respectively, are determined by the small correction $\frac{1}{\nu}\text{Im}[q(s_r)]|s_r|^{-\nu}$ in (3). Eqs. (3) and (4) clearly demonstrate us that only for an asymptotically nonlinear trajectories (5) the resonance width is proportional to its mass, i.e. $\Gamma_r = -2\,M_r \tan\left(\frac{\phi_r}{2}\right)$ and $\phi_r \to \pi\left(1 - \frac{1}{\nu}\right)$ in the limit $|s_r| \to \infty$. Contrarily, for the linear trajectory $\nu = 1$ the resonance width behaves as

$$\Gamma_r = 2\,\text{Im}[q(s_r)]|s_r|^{-\frac{1}{2}} \sim \frac{2\,\text{Im}[q(s_r)]}{|s_r|} M_r\,, \tag{6}$$

where in the last step we used expression (4) for the resonance mass. Since the function $q(s)$ is a small correction to the linear $s$-dependence, from the right hand side equation (6) one concludes that for asymptotically linear trajectories the width of heavy resonances cannot be proportional to their mass since the ratio $\frac{\text{Im}[q(s_r)]}{|s_r|} \to 0$ for large $|s_r|$. Thus, we obtain a very important conclusion that only the asymptotically nonlinear Regge trajectories (2) with $\nu$ lying between $\frac{1}{2}$ and 1 lead to the linear mass dependence of resonance width, i.e. $\Gamma_r \sim M_r$, whereas asymptotically linear Regge trajectories (2) with $\nu = 1$ generate weaker mass dependence of the width. However, this general analysis cannot determine either the form of function $q(s)$ or the range of $s$ at which such an asymptotic behavior is valid.

Fortunately, both of these questions can be answered within the FWM [9]. Thus, the FWM tells us that for excited resonances that belong to the continuous part of the mass-volume spectrum of QG bags the width dependence (6) starts to develop already for $M_r \geq M_0 \approx 2.5$ GeV. Also it predicts that $q(s) \sim s^{\frac{3}{4}}$ which leads to $\Gamma_r \sim \sqrt{M_r}$. Such a conclusion gives a natural explanation of the observed huge deficit [19] in the number of hadronic resonances compared to the statistical bootstrap model [11]. Using these results we conclude that the linear mass dependence of the resonance width assumed in [18] cannot not be used to model the decay of heavy QG bags. Also it can be that application of the obtained results to the decay of Hagedorn states [14, 15] can give quantitatively different outcome.

## 3 Constructing the Data Sets

The FWM predicts an existence of a universal Regge trajectory for heavy QG bags. However, a determination of the parameters of such a trajectory immediately faces two principal difficulties.

The first of these difficulties is that the universal trajectory corresponds to heavy (excited) resonances with mass above $M_0 \approx 2.5$ GeV [9], while the experimental data in this region are absent. The second one is related to the fact that usual fitting procedure is not suited to this task.

For an analysis we choose the best studied trajectories [5, 20] of $\rho_{J^{--}}$, $a_{J^{++}}$ mesons of isospin 1, and of $\omega_{J^{--}}$, $f_{J^{++}}$ mesons of isospin 0 with $J \leq 6$. Nowadays there are three mesons in each of these trajectories [21]. These mesons are well suited to our purpose since, firstly, the parameters of their trajectories are close to each other [5, 20], and, secondly, the masses of $a_{6^{++}}$ and $f_{6^{++}}$ mesons are $2.45 \pm 0.13$ GeV and $2.465 \pm 0.05$ GeV, respectively, i.e. their masses are close to the value of $M_0$. Since here we would like to simultaneously fit the masses and the widths of resonances, we restrict ourselves to an analysis of $\rho_{J^{--}}$, $a_{J^{++}}$, $\omega_{J^{--}}$ and $f_{J^{++}}$ mesons because among other hadrons consisting of $u$- and $d$-quarks only the light hadronic resonances are well studied compared to these mesons. Also here we do not examine the hadrons representing the heavier quark flavors, since they require a separate analysis.

Although the trajectories of $\rho_{J^{--}}$, $a_{J^{++}}$, $\omega_{J^{--}}$ and $f_{J^{++}}$ mesons are similar, they are not identical. Since there is no a priori knowledge on what trajectory is more close to the asymptotic one, we cannot reject any of the data points by claiming that one of the data set is wrong. Also we cannot simply fit all trajectories by the same set of parameters since in many cases the masses and width of the mesons with same value of resonance spin $J$ are rather different and their error bars even do not overlap. Therefore, our first task is to determine the correct set of data to be fitted with the corresponding errors.

The non-zero difference of meson masses of the same spin is a reflection of the chiral symmetry breaking in the confined phase. It is expected that for excited mesons the effect of chiral symmetry breaking gets weaker [22, 23]. This expectation is in line with the FWM prediction of the universal trajectory existence. However, from the practical point of view it is necessary to account the effect of chiral symmetry breaking into the fitting procedure. Clearly, the natural measure of the chiral symmetry breaking, which has to be included into the fitting, is the mass difference of $\delta M_J^o = |M_{\omega_J} - M_{\rho_J}|$ for odd values of $J$ and $\delta M_J^e = |M_{f_J} - M_{a_J}|$ for even $J$ values. Since the mass differences $\delta M_J^o$ and $\delta M_J^e$ are much smaller than the masses of corresponding mesons, i.e. the chiral symmetry breaking effect is small, then it seems reasonably to expect that the universal trajectory should be located very close to or inside of the mass interval of mesons having the same spin. The same, of course, should be true for the resonance width. In addition it is necessary to account for the experimental errors of resonance masses and widths.

Therefore, the **hypothesis A** to be verified by the fit of experimental data is as follows: for the spin $J$ the mass and width defined by the universal trajectory are located within the interval

$$M_r^{exp} \in \left[\min\{M_J^{min} - \delta M_J^{min}; M_J^{max} - \delta M_J^{max}\}; \ \max\{M_J^{min} + \delta M_J^{min}; M_J^{max} + \delta M_J^{max}\}\right], \quad (7)$$

$$\Gamma_r^{exp} \in \left[\min\{\Gamma_J^{min} - \delta \Gamma_J^{min}; \Gamma_J^{max} - \delta \Gamma_J^{max}\}; \ \max\{\Gamma_J^{min} + \delta \Gamma_J^{min}; \Gamma_J^{max} + \delta \Gamma_J^{max}\}\right], \quad (8)$$

respectively. Here $M_J^{min}$ and $\delta M_J^{min}$ ($M_J^{max}$ and $\delta M_J^{max}$) denotes the minimal (maximal) value of meson mass of spin $J$ and its experimental error, respectively. In other words, for each $J$ the mass (width) of the universal Regge trajectory is assumed to be located inside the widest interval that can be constructed from masses (widths) of two mesons and their experimental errors. Thus, our fitting hypothesis **A** relies on the maximal uncertainty in the experimental mass and width values, which, on the one hand, allows us to account for the experimental splitting in the masses and widths of resonances, and, on the other hand, to reduce an individual influence of each of four trajectories analyzed. Such an assumption allows us to determine the mean values of mass and width to be fitted for each $J$ as

$$M_r^{exp} \equiv \frac{1}{2} \left[\min\{M_J^{min} - \delta M_J^{min}; M_J^{max} - \delta M_J^{max}\} + \max\{M_J^{min} + \delta M_J^{min}; M_J^{max} + \delta M_J^{max}\}\right], \quad (9)$$

|  | $A_{exp}$ | $A_{fit}$ | $B_{exp}$ | $B_{fit}$ | $C_{exp}$ | $C_{fit}$ |
|---|---|---|---|---|---|---|
| $M_1 \pm \delta M_1$ | $0.7754 \pm 0.00734$ | $0.7749$ | $0.7758$ | $0.7619$ | $0.7690 \pm 0.0009$ | $0.76897$ |
| $\Gamma_1 \pm \delta\Gamma_1$ | $0.0792 \pm 0.0708$ | $0.0342$ | $0.0788$ | $0.0510$ | $0.1490 \pm 0.001$ | $0.14895$ |
| $M_2 \pm \delta M_2$ | $1.2971 \pm 0.0233$ | $1.3087$ | $1.2970$ | $1.2936$ | — | — |
| $\Gamma_2 \pm \delta\Gamma_2$ | $0.1451 \pm 0.0423$ | $0.1286$ | $0.1445$ | $0.1704$ | — | — |
| $M_3 \pm \delta M_3$ | $1.6770 \pm 0.014$ | $1.6893$ | $1.6779$ | $1.6748$ | $1.6888 \pm 0.0021$ | $1.6891$ |
| $\Gamma_3 \pm \delta\Gamma_3$ | $0.1645 \pm 0.0135$ | $0.1743$ | $0.1645$ | $0.2297$ | $0.1610 \pm 0.01$ | $0.1638$ |
| $M_4 \pm \delta M_4$ | $2.0101 \pm 0.0191$ | $2.0028$ | $2.0095$ | $1.9895$ | — | — |
| $\Gamma_4 \pm \delta\Gamma_4$ | $0.2820 \pm 0.062$ | $0.2049$ | $0.2750$ | $0.2700$ | — | — |
| $M_5 \pm \delta M_5$ | $2.2725 \pm 0.0925$ | $2.2753$ | $2.2900$ | $2.2632$ | $2.3300 \pm 0.035$ | $2.27$ |
| $\Gamma_5 \pm \delta\Gamma_5$ | $0.3625 \pm 0.1375$ | $0.2282$ | $0.3600$ | $0.3010$ | $0.4000 \pm 0.1$ | $0.177$ |
| $M_6 \pm \delta M_6$ | $2.4500 \pm 0.13$ | $2.5171$ | $2.4575$ | $2.5056$ | — | — |
| $\Gamma_6 \pm \delta\Gamma_6$ | $0.4000 \pm 0.25$ | $0.2489$ | $0.3275$ | $0.3285$ | — | — |
| $\chi^2$ | — | $0.7739$ | — | $1.3109$ | — | $4.$ |

**Table 1.** The masses and widths (both given in GeV) of the data sets **A**, **B** and **C** along with the results obtained by their fitting (see more details in the text). The column $A_{exp}$ contains the data points and their errors for the hypothesis **A**. In the column $B_{exp}$ there data points corresponding to the hypothesis **B**, while they have the same error $\Delta$. The column $C_{exp}$ consists of the experimental data on masses and widths of $\rho_{1--}$, $\rho_{3--}$ and $\rho_{5--}$ mesons which are given to demonstrate the typical results of the four parametric fit. The last row contains the corresponding $\chi^2$ per number of degrees of freedom to show the quality of the fit.

$$\Gamma_r^{exp} \equiv \frac{1}{2}\left[\min\{\Gamma_J^{min} - \delta\Gamma_J^{min}; \Gamma_J^{max} - \delta\Gamma_J^{max}\} + \max\{\Gamma_J^{min} + \delta\Gamma_J^{min}; \Gamma_J^{max} + \delta\Gamma_J^{max}\}\right], \quad (10)$$

and their errors as

$$\delta M_r \equiv \frac{1}{2}\left[\max\{M_J^{min} + \delta M_J^{min}; M_J^{max} + \delta M_J^{max}\} - \min\{M_J^{min} - \delta M_J^{min}; M_J^{max} - \delta M_J^{max}\}\right], (11)$$

$$\delta\Gamma_r \equiv \frac{1}{2}\left[\max\{\Gamma_J^{min} + \delta\Gamma_J^{min}; \Gamma_J^{max} + \delta\Gamma_J^{max}\} - \min\{\Gamma_J^{min} - \delta\Gamma_J^{min}; \Gamma_J^{max} - \delta\Gamma_J^{max}\}\right]. \quad (12)$$

The set **A** defined by Eqs. (9)–(12) from the experimental data is shown in the second column of Table 1. As one can see from this column the errors of $J = 1$ mass, $J = 3$ mass and $J = 3$ width are essentially smaller than other errors, and, as we will see in the next section, they essentially affect the results of fitting. Therefore, in order to weaken such a dependence we would like to verify the **hypothesis B** that for spin $J$ the universal Regge trajectory passes through the corridor $\pm \Delta$ taken from the arithmetical average of masses and width of corresponding mesons (set **B**). Here $\Delta$ is some typical value of experimental uncertainties, which in the actual minimization procedure was chosen $\Delta = 0.035$ GeV. It is, however, clear that scaling of $\Delta$ does not change the location of a $\chi^2$ minimum in the space of parameters, although it changes the value of mean deviation squared per number of degrees of freedom $\chi^2$ and the error bars of the fitting parameters.

Comparing the sets **A** and **B** (see columns $A_{exp}$ and $B_{exp}$ in Table 1) one can see that the corresponding mass and width values, except for $\Gamma_6$, are very close to each set of data. The difference in the uncertainties of these sets allows us to study the stability of the fitting parameters.

## 4 Fitting Procedure and Results

For our analysis we choose the simplest parameterization of the Regge trajectory which satisfies the requirements (I) - (III) and obeys the FWM predictions [9] that $q(s) \sim s^{\frac{3}{4}}$:

$$\alpha(s) = \alpha_0 + g_R^2 \left[ s + A_R(-s)^{\frac{3}{4}} - iB_R \right]. \quad (13)$$

Note that the term $(-s)^{\frac{3}{4}}$ in (13) has a correct behavior in the complex $s$-plane [8]. Two additional parameters to the asymptotical linear trajectory (2), $\alpha_0$ and $B_R$, define the real and imaginary parts of $\alpha(0)$ at $s = 0$, respectively, i.e. $Re(\alpha(0)) \equiv \alpha_0$ and $Im(\alpha(0)) \equiv -g_R^2 B_R$. The constant $A_R$ defines the phase $\phi_r$ of a resonance in the complex energy plane as

$$\sin(\phi_r) = A_R \sin\left(\tfrac{3}{4}(\pi - \phi_r)\right) \sqrt{\tfrac{\cos\left(\tfrac{\phi_r}{2}\right)}{M_r}} + \tfrac{B_R}{M_r^2} \cos^2\left(\tfrac{\phi_r}{2}\right). \quad (14)$$

Clearly, this equation is an explicit form of Eq. (3) for the trajectory (13). As we discussed in Section 2, $-\pi < \phi_r < 0$, which leads to the inequality $B_R < -A_R \sin\left(\tfrac{3}{4}(\pi - \phi_r)\right) \left[M_r^2 + \tfrac{\Gamma_r^2}{4}\right]^{\frac{3}{4}}$ that should hold for masses, widths and phases of all resonances belonging to the trajectory (13).

In fact, we used more complicated parameterizations of the trajectory $\alpha(s)$ then that one of Eq. (13), but they did not give any improvement of the fit. In particular, in order to avoid the singularity of the intercept $\tfrac{d\,\alpha(s)}{d\,s}$ at $s = 0$, instead of the term $(-s)^{\frac{3}{4}}$ in (13) we also considered $(-(s + C_0))^{\frac{3}{4}}$ with a complex constant $C_0$. However, this modification increases the overall value of $\chi^2$ per number of degree of freedom since the reduction in the number of degree of freedom for one or two units has a dominant effect.

The spin of the resonance at its position in the complex $s$-plane $s_r = |s_r| e^{i\phi_r}$ is given by the expression

$$J_r = Re\left(\alpha_R(s_r)\right) = \alpha_0 + g_R^2 M_r^2 \frac{\left[\sin\left(\tfrac{1}{4}(3\pi + \phi_r)\right) - \tfrac{B_R}{M_r^2} \cos^2\left(\tfrac{\phi_r}{2}\right) \cos\left(\tfrac{3}{4}(\pi - \phi_r)\right)\right]}{\cos^2\left(\tfrac{\phi_r}{2}\right) \sin\left(\tfrac{3}{4}(\pi - \phi_r)\right)}, \quad (15)$$

whereas its mass $M_r$ and width $\Gamma_r$ are defined by Eqs. (4). As one can see from Eqs. (14) and (15) the parameter $B_R$ enters in these equations only in combination $\tfrac{B_R}{M_r^2}$. This fact clearly demonstrates an importance of $B_R$ parameter for small values of resonance mass, while for large values of $M_r$ it generates a small correction to an asymptotic behavior of the trajectory.

Eq. (15) can be rewritten in a form

$$M_r = \frac{1}{g_R} \left[ \frac{(J_r - \alpha_0) \cos^2\left(\tfrac{\phi_r}{2}\right) \sin\left(\tfrac{3}{4}(\pi - \phi_r)\right)}{\sin\left(\tfrac{1}{4}(3\pi + \phi_r)\right) - g_R^2 B_R \cos^2\left(\tfrac{\phi_r}{2}\right) \cos\left(\tfrac{3}{4}(\pi - \phi_r)\right)} \right]^{\frac{1}{2}} \quad (16)$$

which is more convenient for finding out the resonance mass for known spin $J_r$ and phase $\phi_r$. The advantage of Eq. (16) is that its right hand side does not depend on $M_r$. This allows us to simplify the searches for the minimum of $\chi^2$-function

$$\chi^2(A_R, B_R, g_R, \alpha_0) = \frac{1}{N_{dof}} \sum_r \left[ \frac{[M_r - M_r^{exp}]^2}{\delta M_r^2} + \frac{[\Gamma_r^{exp} + 2M_r \tan\left(\tfrac{\phi_r}{2}\right)]^2}{\delta \Gamma_r^2} \right], \quad (17)$$

by solving numerically the system of Eqs. (14) and (16) for two unknown variables $\phi_r$ and $M_r$ of a resonance having spin $J_r$ for a given set of $A_R, B_R, g_R$ and $\alpha_0$ values. Here $N_{dof}$ denotes the number of independent degrees of freedom in the fitting, $M_r^{exp}$ and $\Gamma_r^{exp}$ are, respectively, the mass and width of the resonance of spin $J_r$ taken from the data sets defined in a preceding section, whereas $\delta M_r$ and $\delta \Gamma_r$ are the corresponding uncertainties. The minimum of the $\chi^2$-function (17) was found by independent variation of the fitting parameters $A_R, B_R, g_R$ and $\alpha_0$.

| Parameter | $A_{fit}$ | $B_{fit}$ | $C_{fit}$ |
|---|---|---|---|
| $\alpha_0$ (GeV)$^0$ | $0.4260 \pm 0.0120$ | $0.4250 \pm 0.0180$ | $0.449 \pm 0.007$ |
| $g_R$ (GeV)$^{-1}$ | $0.8815 \pm 0.0049$ | $0.8667 \pm 0.0155$ | $0.906 \pm 0.006$ |
| $A_R$ (GeV)$^{\frac{1}{2}}$ | $-0.287 \pm 0.0110$ | $-0.377 \pm 0.0218$ | $-0.157 \pm 0.008$ |
| $B_R$ (GeV)$^2$ | $0.1033 \pm 0.0504$ | $0.1327 \pm 0.0119$ | $-0.050 \pm 0.007$ |

**Table 2.** The parameters of the asymptotic Regge trajectory (13) obtained by the fitting of the sets **A**, **B** and **C** given in Table 1. The errors correspond to a standard (one $\sigma$) deviations from the data point values of the corresponding set.

It is necessary to stress here that such a procedure provides an exact treatment of the resonance width in contrast to a popular approximate relation [7]

$$\Gamma_r \approx \frac{Im(\alpha(M_r^2))}{M_r\, Re\left(\left.\frac{d\alpha(s)}{d\,s}\right|_{s=M_r^2}\right)}, \quad (18)$$

which can be used for very heavy resonances only, while for $\rho_{1^{--}}$ and $\omega_{1^{--}}$ it deviates from an exact result by 20–30 %.

We performed the four parametric fit of the data set **A** defined by Eqs. (9)–(12). The results are given in Tables 1 and 2 and shown in Figs. 1–3. Although the $\chi_A^2 \approx 0.774$ value is smaller than 1, the close inspection shows that, in contrast to an excellent fit of resonance masses, the fit of their widths seems not very satisfactory. From Figs. 2 and 3 one can clearly see that the set **A** data for the width are perfectly described for $J_r \leq 3$ only, whereas the obtained width of resonances $\Gamma_r$ with $J_r > 3$ formally provides the minimum of $\chi^2$-function because $\Gamma_r \in [\Gamma_r^{exp} - \delta\Gamma_r; \Gamma_r^{exp}]$, but the behavior of $\Gamma_r$ does not reproduce the trend of the data for $J_r > 3$. The reason for such a behavior is that the mass and width uncertainties of set **A** are essentially smaller for the resonances of spin $J_r \leq 3$ than for that ones of higher spin values. Exactly the same behavior we obtained while examining the individual trajectories of $\rho_{J^{--}}$, $\omega_{J^{--}}$, $a_{J^{++}}$ and $f_{J^{++}}$ mesons. The last column of Table 1 shows the results of the $\rho_{J^{--}}$ trajectory fit. One can see that the masses and widths of $\rho_{1^{--}}$ and $\rho_{3^{--}}$ mesons are reproduced perfectly whereas the mass of $\rho_{5^{--}}$ meson is almost two standard deviations off from its experimental mean value and its width is about two and half standard deviations off the mean experimental value of a width. An evident origin for such outcome of the fit is rooted in very small experimental errors of $\rho_{1^{--}}$ and $\rho_{3^{--}}$ mesons compared to the errors of $\rho_{5^{--}}$ meson. Clearly, the large value of $\chi_C^2 \approx 4$ for the $\rho_{J^{--}}$ trajectory is generated by the $\rho_{5^{--}}$ meson data points.

After realizing this fact we examined the stability of the obtained results. For this purpose we formulated the hypothesis **B** and analyzed it. Comparing the data sets **A** and **B**, from Table 1 it is clearly seen that the main difference between them is due to the value of errors: in contrast to the set **A**, all errors for the set **B** are chosen to be equal to $\Delta$ (democratic choice). The results of the set **B** fitting are given in Tables 1 and 2 and shown in Figs. 1–3. From Table 1 one can see that the obtained fit corresponds much better to the data trend of set **B**. In fact, there are only two data points, $\Gamma_3$ and $\Gamma_5$, which are about 60 MeV off the corresponding experimental values. Such a result seems to be a remarkable success for the trajectory (13) which is expected to be valid in the limit $|s| \to \infty$.

From Table 2 it is seen that the most sensitive parameter to the change of data sets is $A_R$, whereas $\alpha_0$ and $g_R$ are almost insensitive parameters and $B_R$ demonstrates a moderate sensitivity.

Also from this table it is clear that the values of the fitting parameters $\alpha_0$ and $g_R$ are in good agreement with the corresponding parameters obtained by other groups [5, 20, 22].

Now it is necessary to chose the best fit. Although for the common error $\Delta = 35$ MeV the found value of $\chi_B^2(35\text{MeV}) \approx 1.3109$ for the set **B** is slightly larger than that one for the set **A**, one cannot simply favor the fit **A** on the basis of smaller $\chi^2$ value. The problem is that one can increase $\Delta$ above the critical value $\Delta_c = 35\,\text{MeV}\left[\frac{\chi_B^2(35\text{MeV})}{\chi_A^2}\right]^{\frac{1}{2}} \approx 45.55$ MeV and in this way it is possible to reduce $\chi_B^2(\Delta > \Delta_c)$ below $\chi_A^2$. Clearly, the re-scale of the common error $\Delta$ would not change the values of fitting parameters and the minimum location for the set **B**. Therefore, it seems that the hypothesis **B** with the common error $\Delta > \Delta_c$ is more probable than the hypothesis **A**, but rather small value of $\chi_A^2$ does not allow us to simply reject it. Thus, we need some additional criterion to favor one of these hypotheses.

Such a criterion is provided by the FWM which predicts the asymptotic behavior of the width of large/heavy QG bags on the basis of the lattice QCD data. Indeed, the FDM [9] allows one to extract the asymptotic mass dependence of the QG bag width $\Gamma_R$ of the mean volume $V_r = \frac{V_0}{M_0} M_r$ and mass $M_r$

$$\Gamma_R(M_r) = 2\,C_\gamma \sqrt{\frac{2\,\ln(2)\,T_{co}^5\,V_0}{M_0} M_r}\,. \tag{19}$$

from a variety of the lattice QCD data [24, 25, 26] for vanishing baryonic density. Here $V_0 = 1$ fm$^3$ is the minimal volume of large QG bags, $T_{co}$ is the cross-over temperature at zero baryonic density, and the constant $C_\gamma$ weakly depends on the number of elementary degrees of freedom in the analyzed lattice QCD model [9]: $C_\gamma \approx 1.28$ corresponds to the pure gluodynamics for the $SU(2)_C$ color group [24], $C_\gamma \approx 1.22$ describes the $SU(3)_C$ color group lattice QCD data with two quark flavors [25] and the value $C_\gamma \approx 1.3$ corresponds to the recent lattice QCD data for the $SU(3)_C$ color group with three quark flavors [26].

Equating the asymptotic form of the width from Eq. (19) with that ones obtained from the fitting of sets **A** and **B**, we can determine the corresponding value of the cross-over temperature and compare it with the established value. Taking the limit of large resonance mass $M_r \to \infty$ in Eq. (14), one finds the asymptotic behavior of the resonance phase and width

$$\phi_r\Big|_{M_r \to \infty} \to A_R \frac{\sin\left(\frac{3}{4}\pi\right)}{\sqrt{M_r}} \to -0\,, \tag{20}$$

$$\Gamma_r\Big|_{M_r \to \infty} \to -M_r\,\phi_r \to -A_R \sqrt{\frac{M_r}{2}}\,. \tag{21}$$

From Eqs. (19) and (21) we find out the following expressions:

$$A_R = -4\,C_\gamma\,T_{co}^{\frac{5}{2}} \sqrt{\frac{\ln(2)\,V_0}{M_0}} \quad \Leftrightarrow \quad T_{co} = \left[\frac{A_R^2\,M_0}{\ln(2)(4\,C_\gamma)^2\,V_0}\right]^{\frac{1}{5}}. \tag{22}$$

Taking the minimal and maximal values of $C_\gamma$ and using the $A_R$ values from Table 2, we determine the possible values of the cross-over temperature for each set: $T_{co}^A \in [153.2; 157.1]$ MeV and $T_{co}^B \in [170.9; 175.3]$ MeV. This result rules out the hypothesis **A**, since its cross-over temperature is lower even than the chemical freeze-out temperature of most abundant hadrons $T_{chem} = 165 \pm 5$ MeV extracted from the nucleus-nucleus collisions at RHIC energy $\sqrt{s_{NN}} = 130$ GeV [27, 28]. On the other hand the cross-over temperature $T_{co}^B$ for the set **B** is in a good agreement with the freeze-out temperature $T_{early} = 170 \pm 5$ MeV of early hadronizing particles for which the kinetic and chemical freeze-out occur simultaneously to the very moment of their hadronization. As we can see from the values of $T_{co}^B$ and $T_{early}$, this is the case for the set **B**. Thus, the data on early freeze-out temperature favor the hypothesis **B**.

The early freeze-out temperature was for the first time established for $J/\psi$-, $\psi'$-mesons and $\Omega^\pm$-hyperons at SPS laboratory energy $E_{NN}^{lab} = 158$ GeV [29] and for $\phi$-mesons and $\Omega^\pm$-hyperons at RHIC energy $\sqrt{s_{NN}} = 130$ GeV [30]. The same conclusion on the early hardonization phenomenon is supported by the recent analysis of transverse momentum spectra of $J/\psi$-, $\phi$-mesons and $\Omega^\pm$ at the top RHIC energy $\sqrt{s_{NN}} = 200$ GeV [31]. A systematic study of the connection between the early kinetic freeze-out of these hadrons and their hadronization process has begun from the work [32] (see also the recent review article [33]).

Also we would like to draw an attention to the discrepancies between the experimental data on the resonance width and their fits. Their mass dependence is shown in Fig. 3. As one can deduce from Fig. 3 the difference between each data set and its fit has some periodic structure, which is more clearly seen for the set **B**. Indeed, Fig. 3 for the set **B** shows that the width of mesons with $J = 1$ and $J = 5$ slightly exceeds the fit values, the width of $J = 3$ meson is somewhat smaller than the fit, while the width of mesons with $J = 2$, $J = 4$ and $J = 6$ almost matches the width of the asymptotic Regge trajectory. The developed approach would allow us to accurately extract such a fine structure, if the experimental uncertainties were smaller. Therefore, to firmly establish this fine structure of the resonance width compared to the asymptotic Regge trajectory we need much more accurate data for all analyzed mesons. However, if this fine structure, indeed, exists, then the mean width of the mesons of spin $J = 7$ (and mass $M_7 \approx 2.723 \pm \Delta_c$ GeV) should be $\Gamma_7 \approx 0.3 \pm \Delta_c$ GeV instead of the set **B** fit value $\Gamma_7^B \approx 0.36 \pm \Delta_c$ GeV, while the mean width of the meson of spin $J = 8$ (and mass $M_8 \approx 2.923 \pm \Delta_c$ GeV) should match the fit **B** value $\Gamma_8^B \approx 0.384 \pm \Delta_c$ GeV, where for an a priori uncertainty we used the critical value of the common error $\Delta_c$ for the set **B**. Note that the list of non-strange mesons in [21] contains the meson X(2750) of spin $J = 7$, mass $M_7 \approx 2.747 \pm 0.032$ GeV and width $\Gamma_7 \approx 0.195 \pm 0.075$, which are very close to our estimate that includes the fine structure discussed above, but, unfortunately, we cannot make a definite conclusion because other quantum numbers of this meson are unknown.

## 5 Conclusions

Here we analyze the asymptotic behavior of Regge trajectories of non-strange mesons. Using the approach of [8] we show that a circulating belief that the width of heavy hadrons linearly depends on their mass simply contradicts to an existence of asymptotically linear Regge trajectories which is expected to be the case by the open string model [34, 35], the closed string model [34], the anti-de-Sitter conformal field theory [36] and the FWM [10].

We analyze the common data sets for masses and widths of $\rho_{J^{--}}$, $\omega_{J^{--}}$, $a_{J^{++}}$ and $f_{J^{++}}$ mesons for the spin values $J \leq 6$ in order to elucidate the parameters of the asymptotically linear Regge trajectory that is consistent with the FWM predictions. Thus, we verified the hypotheses **A** and **B**, which differently define the data uncertainties.

Since the suggested fitting procedure employs the exact expressions for the resonance mass and width derived from the Regge trajectory in the complex energy plain, we obtained a high quality fit. Thus, it is shown that both data sets, **A** and **B**, can be fitted with rather small value of $\chi^2$ per degree of freedom of about 0.774. Therefore, we used the results of the fit to estimate the cross-over temperature value based on the predictions of the FWM and obtained $T_{co}^A \in [153.2; 157.1]$ MeV for the set **A** and $T_{co}^B \in [170.9; 175.3]$ MeV for the set **B**. As it is argued the cross-over temperature obtained from the set **A** is incompatible with the early freeze-out temperature of $J/\psi$-, $\phi$-, $\psi'$-mesons and $\Omega^\pm$-hyperons found out for the laboratory energies at and above $E_{NN}^{lab} = 158$ GeV, while the cross-over temperature of the set **B** is consistent with the early freeze-out temperature of these particles.

Also a close inspection of the mass dependence of resonance width for the analyzed mesons led us to a conclusion of a possible fine (periodic) structure compared to the asymptotic behavior of the Regge trajectory. This result might be of a great interest for the string models of hadrons. However, to firmly establish an existence of such a structure we need, firstly, a higher accuracy of experimental data and, secondly, a thorough verification of the predictions made for the widths and masses of the resonances of spin 7 and 8. Such experimental research will definitely allow us to make more concrete predictions for the asymptotic properties of heavy hadronic resonances.

**Acknowledgments.** We are thankful to N. I. Glushko and E. S. Martynov for fruitful discussions. Also the important comments of C. Greiner are acknowledged. K.A.B. acknowledges the partial support of the Fundamental Research State Fund of Ukraine, Agreement No F28/335-2009 for the Bilateral project FRSF (Ukraine) – RFBR (Russia). The work of A.S.S. was supported in part by the Russian Foundation for Basic Research, Grant No. 08-02-01003.

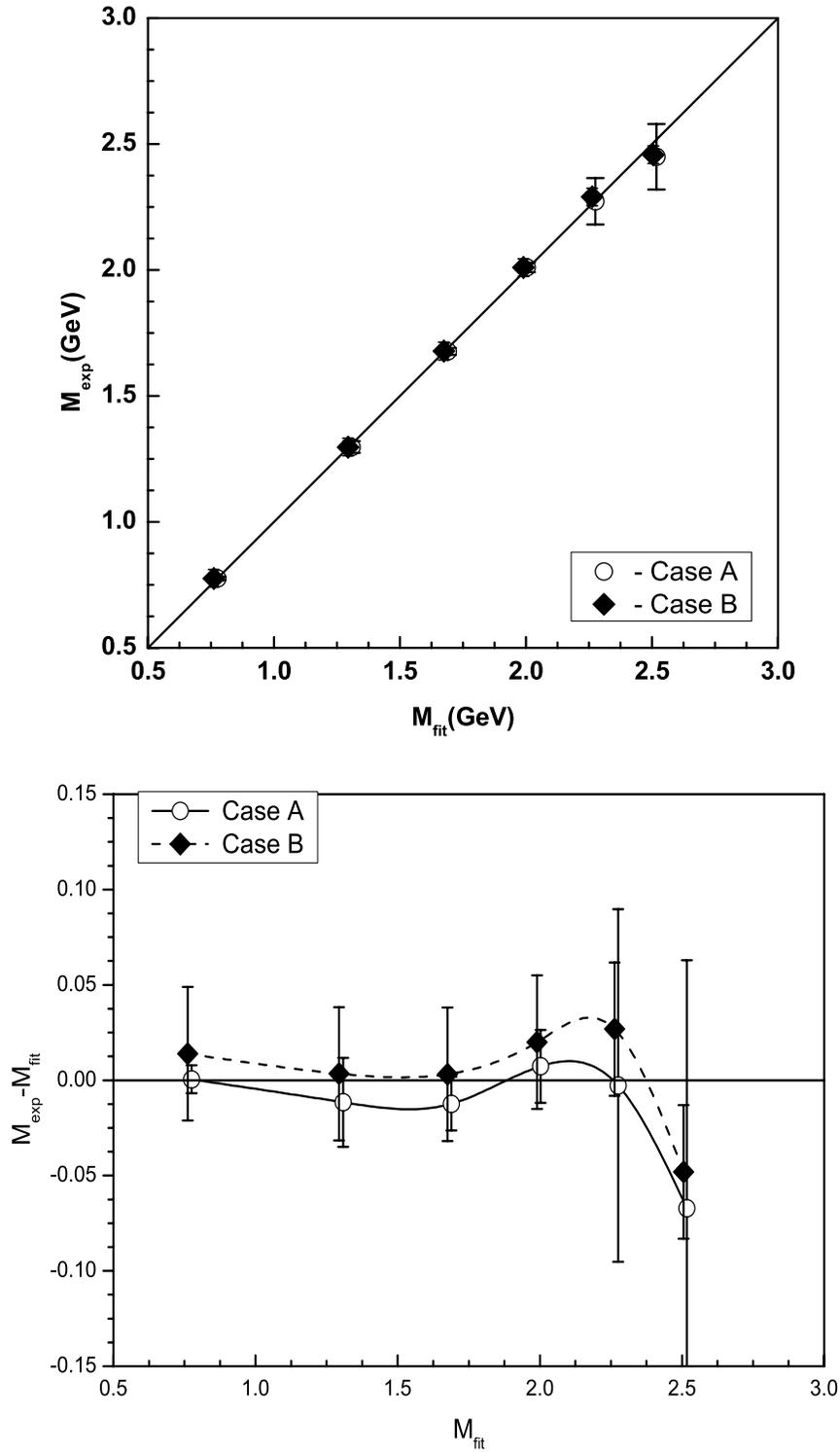

**Fig. 1.** Comparison between the resonance masses of the sets **A**, **B** and their fit by the asymptotic Regge trajectory (13). The upper panel shows the data points as the function of the corresponding mass value of the fit. For most data point the error bars are smaller than the symbols. The lower panel shows the difference of data and fit as the function of the fit value of the resonance mass. The curves in the lower panel are the cubic splines and are shown to guide the eyes.

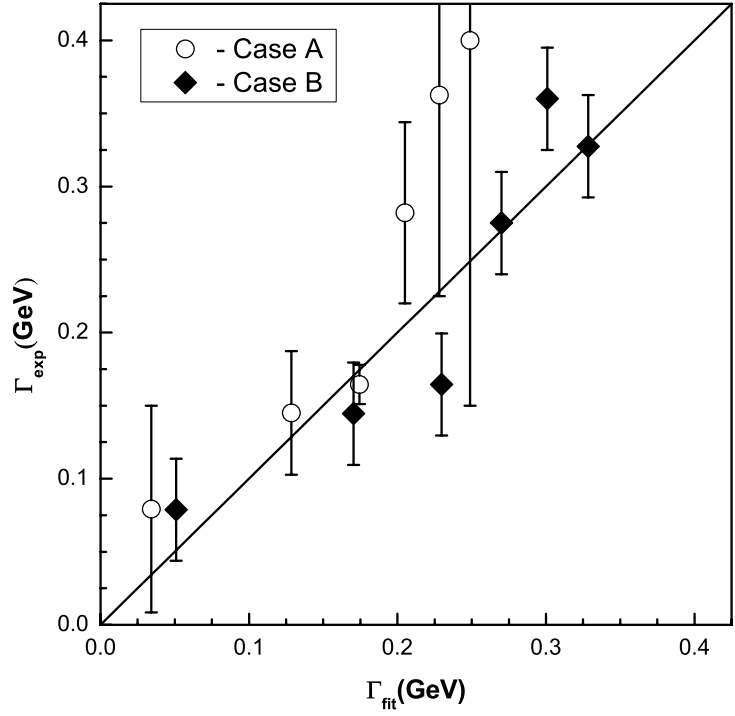
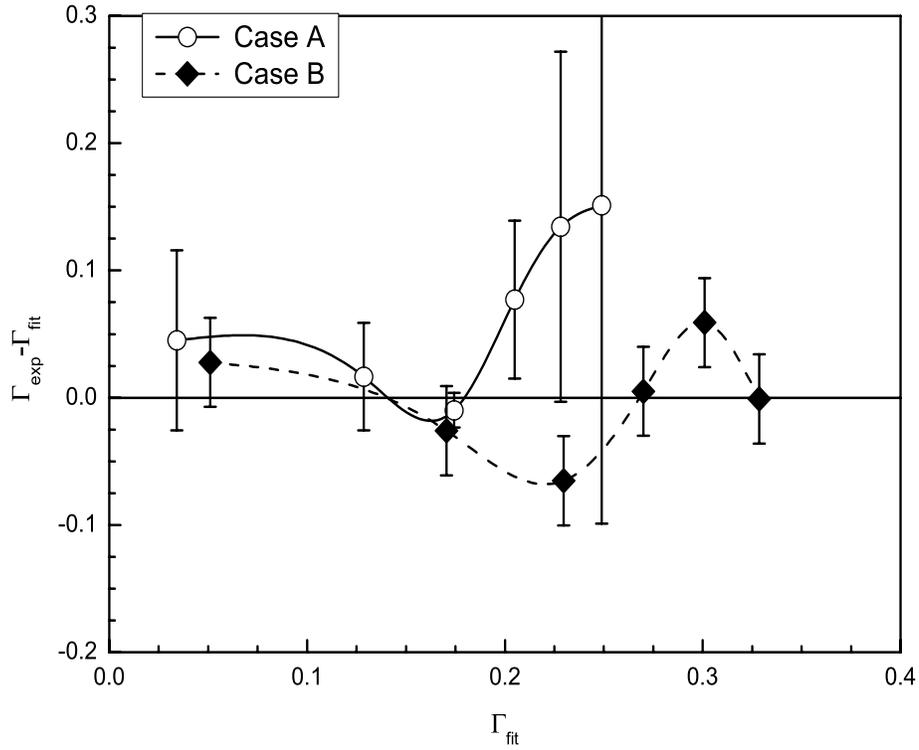

**Fig. 2.** Comparison between the resonance widths of the sets **A**, **B** and their fit by the asymptotic Regge trajectory (13). The upper panel shows the data points as the function of the corresponding width value of the fit. The lower panel shows the difference of data and fit as the function of the fit value of the resonance width. The curves in the lower panel are the cubic splines and are shown to guide the eyes.

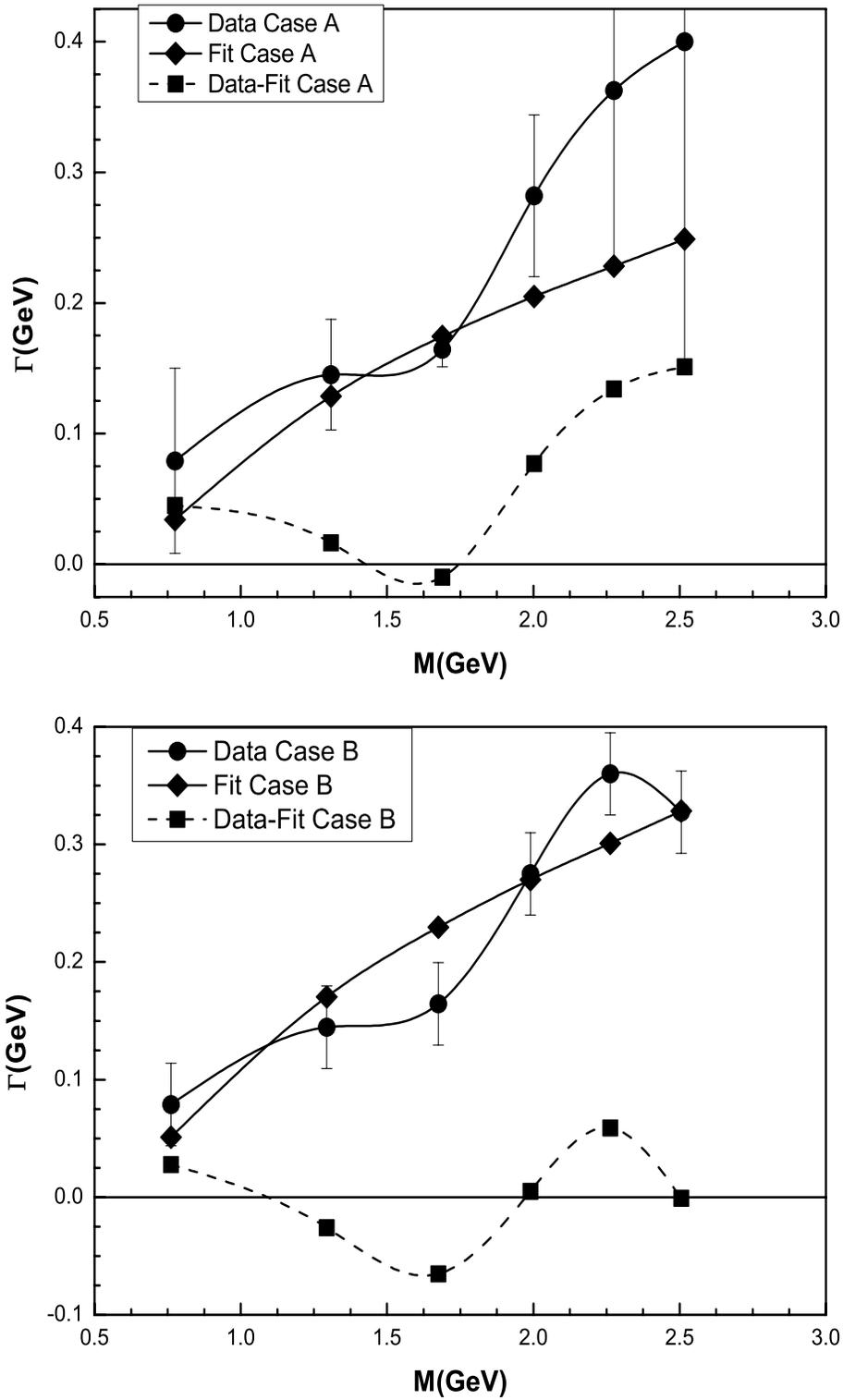

**Fig. 3.** The upper part of each plot shows the resonance width of mesons as the function of the fit value of the resonance mass. The lower part of each plot shows the difference of the resonance width of mesons and the obtained value of the fit as the function of the fit value of the resonance mass. The upper (lower) panel corresponds to the set **A** (**B**). The curves in both panels are the cubic splines and are shown to guide the eyes.